\documentclass[a4paper,11pt]{article}
\usepackage{pos}

\usepackage{csquotes}          
\usepackage{hyperref}          
\usepackage{placeins}          
\usepackage{braket}            

\newcommand{\rb}[1]{\!\left(#1\right)}

\newcommand{\expe}[1]{\text{e}^{#1}}
\DeclareRobustCommand{\eqnr}[1]{Eq.~$\left(\ref{#1}\right)$}

\newcommand{\Fig}[1]{Fig.~\ref{#1}}
\newcommand{\Tab}[1]{Table \ref{#1}}

\newcommand{\Refl}[1]{Ref.~\cite{#1}}    
\newcommand{\Refltwo}[2]{Refs.~\cite{#1,#2}}    

\newcommand{\mpiv}{239(1)}
\newcommand{\Tpc}{T_{\rm pc}}
\newcommand{\mo}{\mathcal{O}}
\newcommand{\gfo}{\gamma_4}

\newcommand{\tr}{\mbox{tr}\,}

\title{Reconstructed (charm) baryon methods at finite temperature on anisotropic lattices}
\onbehalf{on behalf of the \\\textsc{FASTSUM} Collaboration}
\author*[a,e]{Ryan Bignell}
\author[a,b]{Gert Aarts}
\author[a]{Chris Allton}
\author[a]{M.~Naeem Anwar}
\author[a]{Timothy J.~Burns}
\author[c]{Benjamin J\"ager}
\author[d,e]{Jon-Ivar Skullerud}

\affiliation[a]{Department of Physics, Swansea University, Swansea, SA2 8PP, United Kingdom}
\affiliation[b]{European Centre for Theoretical Studies in Nuclear Physics and Related Areas (ECT*) \&
Fondazione Bruno Kessler, Strada delle Tabarelle 286, 38123 Villazzano (TN), Italy}
\affiliation[c]{Quantum Field Theory Center \& Danish IAS, Department of Mathematics and Computer Science, University of Southern Denmark, 5230, Odense M, Denmark}
\affiliation[d]{Department of Theoretical Physics and Hamilton Institute, National University of Ireland Maynooth, County Kildare, Ireland}
\affiliation[e]{School of Mathematics, Trinity College, Dublin, Ireland}

\abstract{Reconstructed-correlator methods have been used to investigate thermal effects in mesonic correlation functions in a fit-independent manner. This technique has recently been extended to the baryonic sector. In this work different ways of implementing this approach for baryon correlators are examined. Using both real and synthetic data it is found that for heavy baryons, such as the $\Xi_{cc}(ccu)$ baryon, different choices are equivalent and that for the lighter nucleon the effect of different implementations is minimal. Further comparison to the so-called \enquote{double ratio} using the \textsc{FASTSUM} Generation 2L thermal ensembles shows that reconstructed-correlator ratios and double ratios 
contain nearly identical quantitative information.
}

\FullConference{The 40th International Symposium on Lattice Field Theory (Lattice 2023)\\
July 31st - August 4th, 2023\\
Fermi National Accelerator Laboratory\\}

\begin{document}
\maketitle

\section{Introduction}

In this contribution we examine the reconstructed-correlator technique~\cite{Ding:2012sp,Kelly:2018hsi,Aarts:2022krz,Aarts:2023nax} applied to baryonic correlation functions. This technique is motivated by the spectral relation for fermionic correlators,
\begin{align}
 \label{eqn:specRel}
      G\rb{\tau; T} = \int_{-\infty}^{\infty}\frac{d\omega}{2\pi}\, K_F\rb{\tau, \omega; T} \rho\rb{\omega; T},
\end{align}
where the kernel $K_F\rb{\tau, \omega; T}$ has a known analytical form and temperature dependence, while the spectral function $\rho\rb{\omega; T}$ is not known and is the item of interest, in particular its temperature dependence. By using the reconstructed correlator, one can examine changes in the spectral function $\rho\rb{\omega; T}$ separate from the changes in the kernel $K_F\rb{\tau, \omega; T}$ as the temperature is changed.

In particular we consider methods which enable the use of this technique for two (fixed-scale) ensembles with different temporal extents $N_{0,1}$ --- and hence different temperatures $T_{0,1}=1/(a_\tau N_{0,1})$, with $a_\tau$ the temporal lattice spacing --- which do not align with the odd factor of $m=N_{0}/N_{1}=T_1/T_0$ required~\cite{Aarts:2023nax}. 
Subsequently we compare this technique to the \enquote{double correlator ratio} method used in \Refltwo{Aarts:2022krz}{Aarts:2023nax} using real lattice QCD data from the \textsc{FASTSUM} thermal ensembles~\cite{Aarts:2020vyb,Aarts:2022krz}. 
The double ratio is defined as
\begin{align}
 \label{eq:dratio}
  R\rb{\tau;T,T_0} &= \left.\frac{G\rb{\tau; T}}{G_{\rm model}\rb{\tau; T, T_0}}\right/\frac{G\rb{\tau; T_0}}{G_{\rm model}\rb{\tau; T_0, T_0}},
\end{align}
where $G\rb{\tau; T}$ is the lattice correlator at temperature $T$ and $G_{\rm model}\rb{\tau; T, T_0}$ a model correlator at temperature $T$ informed by the physics at a lower temperature $T_0$, which in practice is the mass of the ground state at the lowest temperature available. 

\section{Ensemble and correlator details}
\label{sec:LQCD}


We use the anisotropic \enquote{Generation 2L} thermal ensembles of the \textsc{FASTSUM} collaboration~\cite{Aarts:2020vyb,Aarts:2022krz}, consisting of $2+1$ flavours of Wilson fermions and a Symanzik-improved anisotropic gauge action, following the Hadron Spectrum Collaboration~\cite{Edwards:2008ja}.
Full details of the action and parameter values can be found in \Refltwo{Aarts:2020vyb}{Aarts:2022krz}. Ensembles are generated using a fixed-scale approach, such that the temperature is varied by changing $N_{\tau}$, as $T=1/\rb{a_\tau N_{\tau}}$, see  \Tab{tab:ensembles}.

\begin{table}[b]
  \centering
    \caption{\textsc{FASTSUM} Generation 2L ensembles. The lattice size is $32^3 \times N_{\tau}$, with temperature $T = 1/\rb{a_\tau N_{\tau}}$. The spatial lattice spacing is $a_s = 0.11208\rb{31}$ fm, the renormalised anisotropy $\xi = a_s/a_\tau = 3.453(6)$ \cite{Dudek:2012gj,Aarts:2020vyb} and the pion mass $m_\pi = \mpiv$ MeV~\cite{Wilson:2019wfr}. We use $\sim 1000$ configurations and eight (random) sources for a total of $\sim 8000$ measurements at each temperature. The estimate for $\Tpc$ comes from an analysis of the renormalised chiral condensate and equals $\Tpc = 167(2)(1)$ MeV~\cite{Aarts:2020vyb,Aarts:2022krz}.
    }
  \begin{tabular}{r|ccccc||cccccc}
  $N_{\tau}$ & 128 & 64 & 56 & 48 & 40 & 36 & 32 & 28 & 24 & 20 & 16\\ \hline
  $T\,\rb{\text{MeV}}$ & 47 & 95 & 109 & 127 & 152 & 169 & 190 & 217 & 253 & 304 & 380 
\end{tabular}
\label{tab:ensembles}
\end{table}


Baryon correlators are of the form
\begin{align}
  G^{\alpha\alpha^\prime}(x) = \Braket{\mo^\alpha(x)\overline{\mo}^{\alpha^\prime}(0)},
\end{align}
where $\overline{\mo} = \mo^\dagger\gfo$ and $\alpha$, $\alpha^\prime$ are Dirac indices. 
The parity-projected correlation functions at vanishing spatial momentum are \cite{Aarts:2015mma,Aarts:2017rrl,Aarts:2018glk}
$
G_\pm(\tau) = \tr P_\pm G(\tau),
$
where $P_\pm=(1\pm\gamma_4)/2$. These are related as \cite{Aarts:2017rrl}
$
G_\pm(\tau) = -G_\mp(1/T -\tau),
$
implying that the forward- (backward-) propagating states of $G_+(\tau)$ are states with positive (negative) parity.
The three-quark operators used follow \Refltwo{Leinweber:2004it}{Edwards:2004sx} and are described explicitly in \Refl{Aarts:2023nax}. 
In the following we focus upon the $\Xi_{cc}\rb{ccu}$ as an example of a heavy baryon and the nucleon $N\rb{uud}$ as the lightest baryon.


\section{Reconstructed Correlator}

Here we give a brief recap of the method \cite{Aarts:2023nax}. The spectral relation for baryonic correlators is given in Eq.\ (\ref{eqn:specRel}) and the fermionic kernel reads \cite{Aarts:2017rrl}
\begin{align}
  K_F\rb{\tau, \omega; T} = \frac{e^{-\omega\tau}}{1 + e^{-\omega / T}}.
\end{align}
We wish to arrive at an expression relating the correlator at a higher temperature $T$ to one at a lower temperature $T_0$. To do so, we temporarily switch to lattice units, such that  $T=1/N_{\tau}$, $T_0=1/N_{0}$ and $N_{0}/N_{\tau}=m>1$, and introduce the identity, relevant for the fermionic case,
\begin{align}
  1 + \expe{-\omega m N_{\tau}} = \rb{1 + \expe{-\omega N_{\tau}}}\,\rb{\sum_{n=0}^{m-1}\rb{-1}^{n}\,\expe{-n\omega N_{\tau}}}.
\end{align}
This identity requires that $m$ is integer and odd. The fermionic kernel can hence be expressed as
\begin{align}
  K_F(\tau, \omega; &\,  1/N_{\tau}) = \frac{\expe{-\omega\tau}}{1 + \expe{-\omega N_{\tau}}} = \sum_{n=0}^{m-1}\frac{\rb{-1}^{n}\,\expe{-\omega\rb{\tau + n\,N_{\tau}}}}{1 + \expe{-\omega m N_{\tau}}} \quad \nonumber \\
  &= \sum_{n=0}^{m-1}\rb{-1}^{n}\,K_F\rb{\tau + n\,N_{\tau}, \omega; 1/(m N_{\tau})},
  \label{eqn:reconF}
\end{align}
The kernel has therefore been expressed as a summation over a kernel with a longer time extent $m N_{\tau}>N_{\tau}$.

Inserting this resummation into the spectral relation (\ref{eqn:specRel}) relates a correlator at $T=1/N_{\tau}$ to one at a lower temperature $T_0 = 1/N_{0} = 1/\rb{m\,N_{\tau}}$, assuming that the spectral content, i.e.\ $\rho\rb{\omega}$, is unchanged. This yields the reconstructed correlator for fermions
\begin{align}
  G_{\text{rec}}\rb{\tau; 1/N_{\tau}, 1/N_{0}} = \sum_{n=0}^{m-1}\rb{-1}^{n}\,G\rb{\tau + n\,N_{\tau}; 1/N_{0}}.
  \label{eqn:reconG}
\end{align}
If we now switch back to denoting the temperatures with $T$ and $T_0$, the relationship  becomes explicitly
\begin{align}
  G_{\text{rec}}\rb{\tau; T, T_0} = \sum_{n=0}^{m-1}\rb{-1}^{n}\,G\rb{\tau + n/T; T_0}.
  \label{eq:Grecon}
\end{align}

As $m=N_{0}/N_{\tau}=T/T_0$ must be an odd integer, the lattice sizes where this technique can be used are limited in principle. In fact, none of the ensembles in \Tab{tab:ensembles} have this odd integer relation with respect to the ensemble at the lowest temperature ($N_{0} = 128$). 
To get around this limitation, we may consider adding or removing points from the zero-temperature correlator to ensure such an odd-integer ratio. This should be done in such a way to not affect the physics encoded in the correlator.  
One may consider \enquote{padding} the correlator at the minimum of the correlator, with either zeroes or with the minimum value of the correlator. This is done in analogy to the mesonic case \cite{Aarts:2022krz}.
An alternative is removing points from the correlator, again done symmetrically at the minimum of the correlator. In contrast to the padding method, removing points can only be done when $N_{0} > 3N_{\tau}$, which corresponds to $N_{\tau} \le 40$ for our ensembles.

To test these methods, we first apply them to a model correlator
\begin{align}
  G_{\rm model}\rb{\tau; T, T_0} = A_+ K_F\rb{\tau, M^{+}_{0}} + A_-K_{F}\rb{\tau, -M^{-}_{0}} 
  =\frac{A_+\expe{-M^{+}_{0}\tau}}{1 + \expe{-M^{+}_{0}/T}} + \frac{A_-\expe{M^{-}_{0}\tau}}{1 + \expe{M^{-}_{0}/T}},
  \label{eqn:modelG}
\end{align}
where $A_\pm$ is some normalisation chosen to be $1$ and $M^{\pm}_{0}$ are the zero-temperature positive and negative parity masses of the $\Xi_{cc}\rb{ccu}$ and the $N\rb{uud}$ ground states as determined by us in \Refl{Aarts:2023nax} at temperature $T_0=47$ MeV ($N_0=128$).
In units of $a_\tau$, these are
\begin{align}
  a_\tau M^{+}_0\rb{ccu} &= 0.59642(85), \quad &&a_\tau M^{+}_0\rb{uud} = 0.1740(17), \nonumber\\
  a_\tau M^{-}_0\rb{ccu} &= 0.6575(24), \quad &&a_\tau M^{-}_0\rb{uud} = 0.2501(65).
  \label{eqn:masses}
\end{align}
Note that we neglect the uncertainties when constructing the model in \eqnr{eqn:modelG}. The key assumption in this model is that the width of the state is negligible.

Next, we consider the correlators at $N_{\tau}=40$ and $N_{0}=128$. These temporal extents are chosen as they allow us to consider both adding and removing points. Note that for the doubly-charmed baryon the correlator is exponentially suppressed around the centre of the lattice, $G(\tau=20a_\tau)/G(\tau=0)\sim 10^{-5}$, whereas this is less so for the nucleon, with $G(\tau=20a_\tau)/G(\tau=0)\sim 10^{-2}$. Padding with zeroes or with the minimum value are therefore expected to produce quite similar results for the heavier state while somewhat more distinct results can be expected for the lighter nucleon.

To compare the results of the reconstructed correlator, Eq.~(\ref{eq:Grecon}), at $N_\tau=40$ with the actual (model) correlator at the same temperature, we consider the ratio 
\begin{align}
  r\rb{\tau; T, T_0} = G_{\rm rec}\rb{\tau; T, T_0} / G_{\rm model}\rb{\tau; T, T_0}.
  \label{eqn:SRatio}
\end{align}
Note that both correlators in the ratio are evaluated at $N_\tau=40$, but one using the reconstruction method starting from the $N_0=128$ correlator and one using the model parameters $M_0^\pm$ determined at $N_0=128$.

\begin{figure}[t!]
  \centering
  \includegraphics[page=1,width=0.48\columnwidth,keepaspectratio,origin=c]{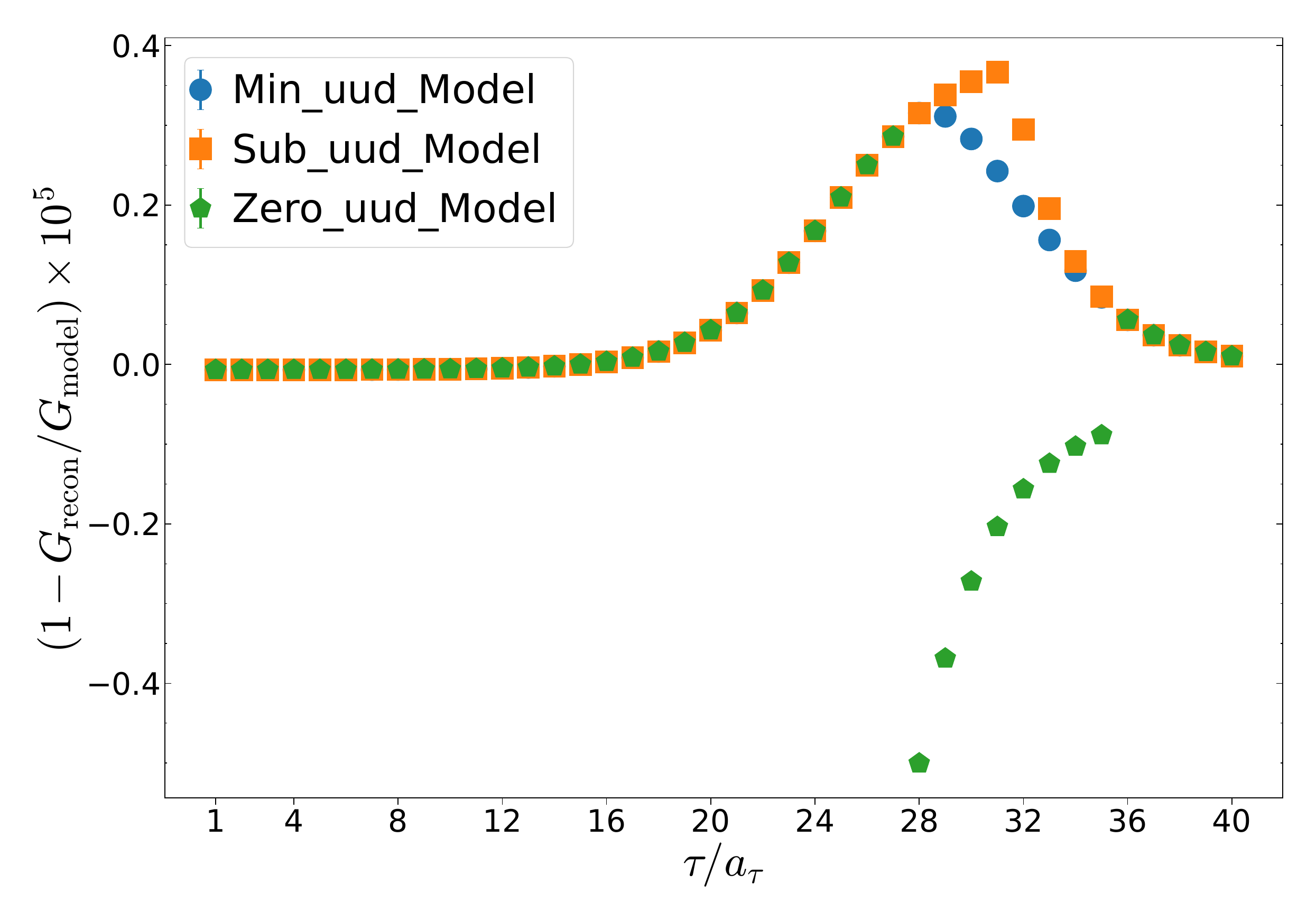}
  \includegraphics[page=1,width=0.48\columnwidth,keepaspectratio,origin=c]{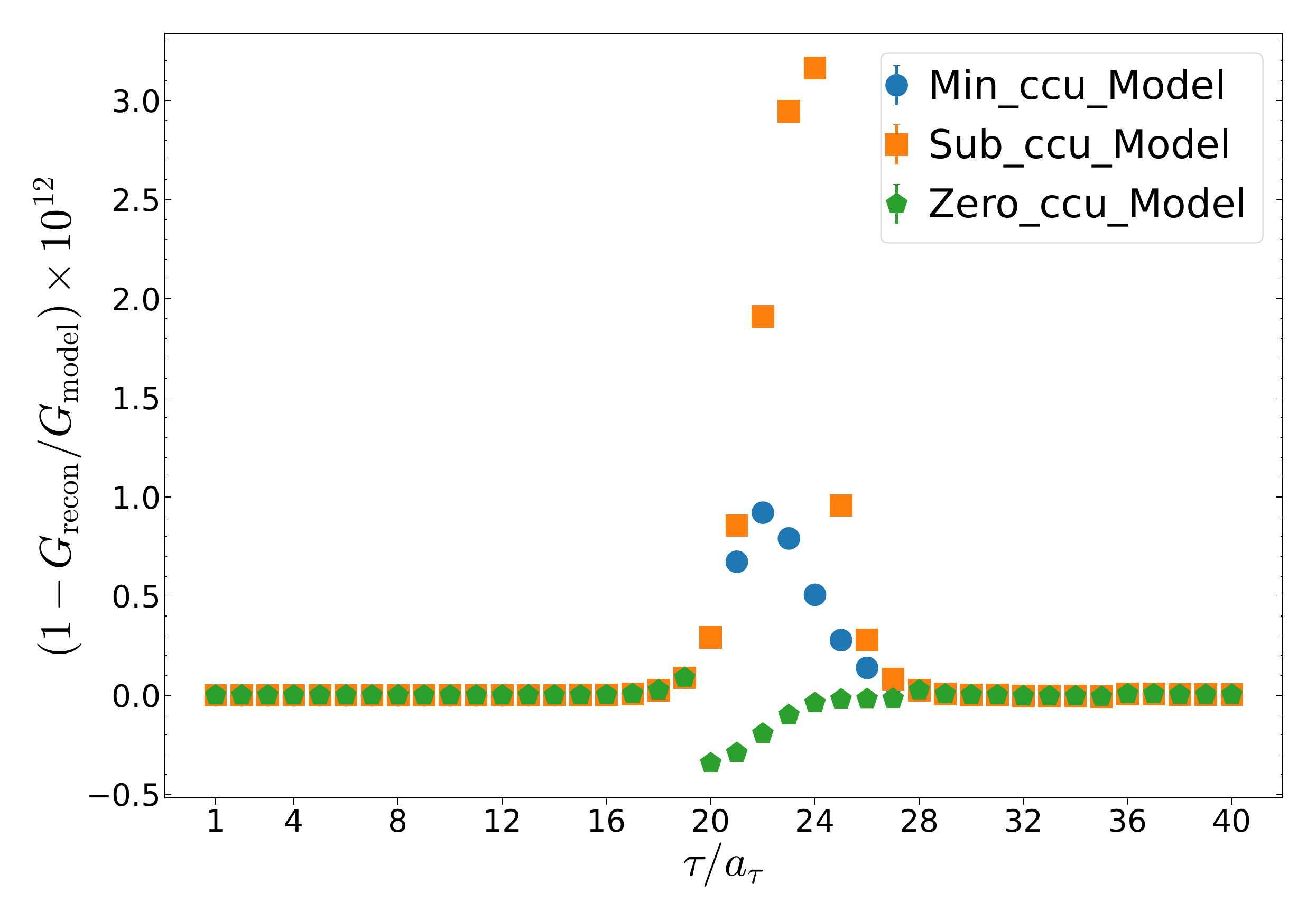}
  \caption{Ratio of reconstructed correlator to model correlator, $(1-G_{\rm rec}/G_{\rm model})\times 10^\alpha$, with $\alpha=5$ for the nucleon (left) and $\alpha=12$ for the $\Xi_{cc}$ (right).
 The three methods (padding with the `Min'imum value, `Sub'traction and padding with `Zero'es) are explained in the main text.
  }
  \label{fig:Correlators}
\end{figure}

In \Fig{fig:Correlators} the ratio is shown using three methods to determine the reconstructed correlator: by padding the $N_0=128$ correlator with the minimal value (Min\_Model), with zeroes  (Zero\_Model), and with points removed (Sub\_Model). Note that what is shown is 
($1-G_{\rm rec}/G_{\rm model})\times 10^\alpha$, with $\alpha=5$ for the nucleon and $\alpha=12$ for the charmed baryon, to  highlight the very small difference between the reconstructed and  model correlators, on the order of $10^{-\alpha}$.
The ``padding with zeroes'' method behaves in an opposite manner to the other two approaches, due to the addition of points with value `0' rather than simply exponentially small.

\begin{figure}[tb]
  \centering
  \includegraphics[width=0.48\columnwidth]{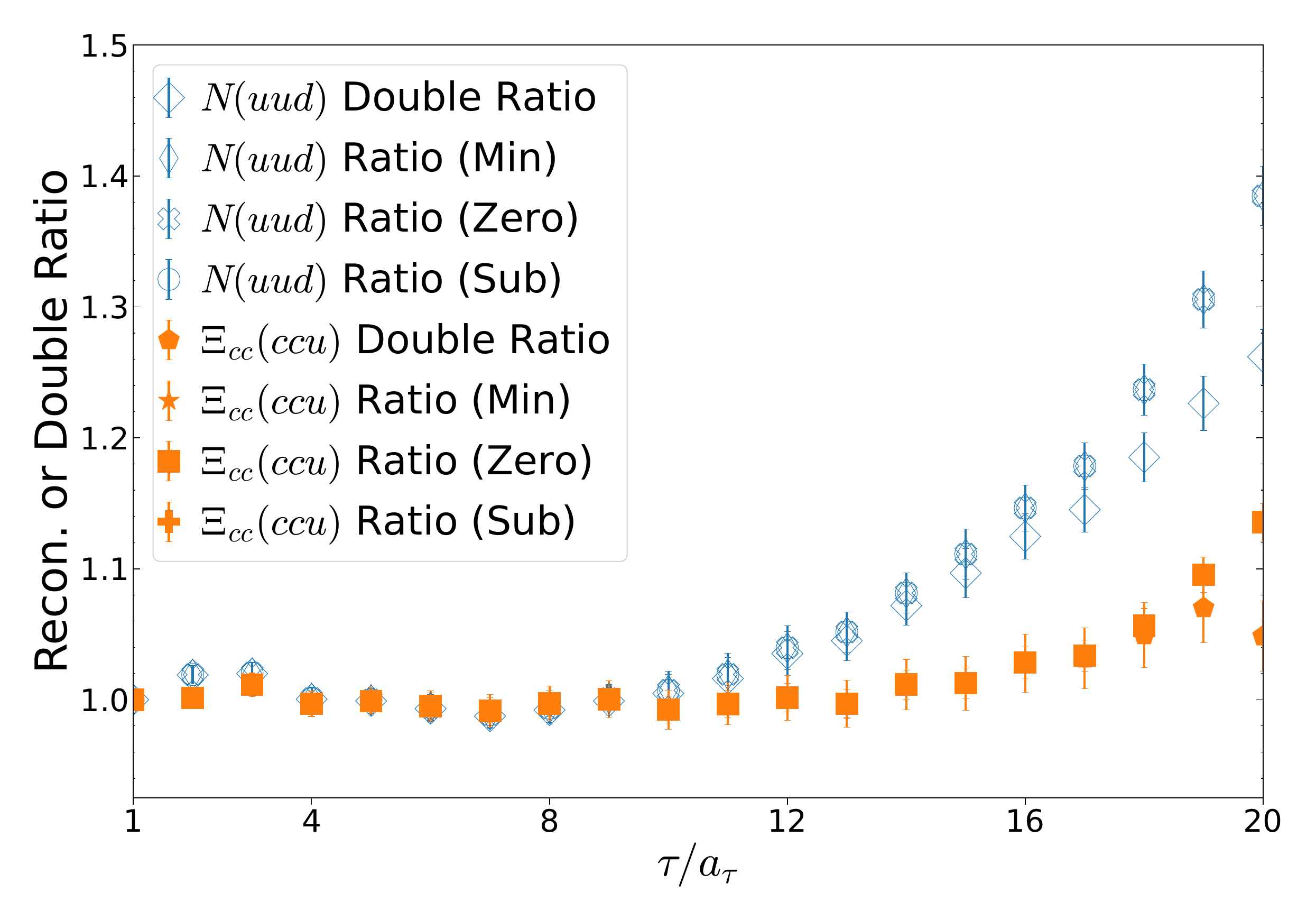}%
  \includegraphics[width=0.48\columnwidth]{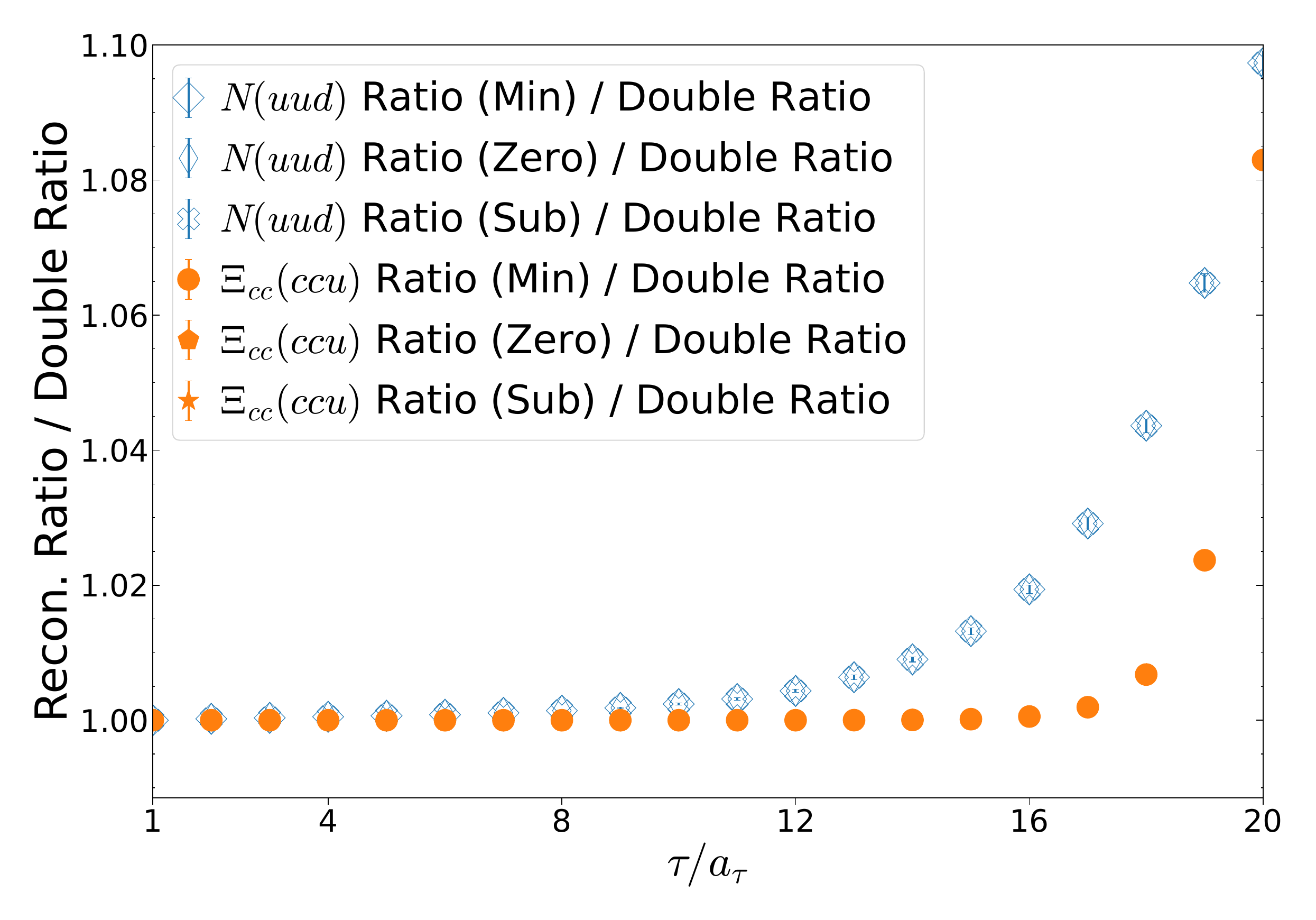}%
  \caption{
  \textbf{Left:} Ratio of reconstructed correlator at $N_{\tau}=40$ to lattice correlator at $N_{\tau}=40$ as well as the double ratio (\ref{eq:dratio}).
  A deviation from 1 shows changes in the spectral content. 
  \textbf{Right:} Ratios of the reconstructed correlator ratio with the double ratio. A deviation from 1 shows differences between the two methods.
  All correlators are normalised to be equal to one at the source.}
  \label{fig:dataRatios}
\end{figure}

\section{Lattice Data}

Following \Refl{Aarts:2023nax}, we now apply these methods to the correlator data from the \textsc{FASTSUM} ensembles,  considering in particular the $\Xi_{cc}(ccu)$ and $N(uud)$ correlators. We compare the reconstructed correlator methods discussed above with the double ratio introduced in \Refltwo{Aarts:2022krz}{Aarts:2023nax} and written in Eq.~(\ref{eq:dratio}). As stated, both the ratio with the reconstructed correlator and the double ratio aim to examine whether changes due to the increase in temperature are present.
When constructing the double ratio, the statistical uncertainty comes from the statistical uncertainty in the mass fit parameters in \eqnr{eqn:masses} as well as from the statistical uncertainty in the correlators.

We explicitly compare these two methods in \Fig{fig:dataRatios} at $N_{\tau}=40$, i.e.\ just below the pseudocritical temperature. The left hand plot shows the reconstructed and double ratios. For the heavier $\Xi_{cc}$, the results agree within uncertainty. For the lighter $N$ there is a visible difference at later times between the double ratio and each of the reconstructed correlator ratios. The qualitative behaviour of interest is, however, still the same.
To further highlight any differences we take ratios of the ratios of correlators shown in the left hand plot. The results are shown in the right hand plot. Here we have removed any uncertainties normally used in the mass parameter which would otherwise obscure any interesting behaviour. This shows that the differences between the three reconstructed correlator methods are indeed very small. 
Note that this test is different from the one in \Fig{fig:Correlators}, as the reconstructed correlator now uses real lattice data as input rather than a model. Finally, the reconstructed and double ratios are in very good agreement as well.


\section{Summary}

In this work, the reconstructed correlator method for baryons was considered. In the \enquote{fixed-scale} approach to thermal lattice QCD, this allows a baryon correlator at a higher temperature to be reconstructed from one at a lower temperature, assuming that the spectral content is unchanged. Hence reconstructed correlators can be used to examine the presence of thermal effects in baryon spectral functions~\cite{Aarts:2023nax}.

To use the technique for two ensembles with temporal extents $N_0$ and $N_\tau$, the ratio $m=N_{0}/N_{\tau}=T/T_0$ should be an odd integer.
As this is usually not the case, some form of alteration of the correlator at the lowest temperature should be performed. Here we considered three approaches: 
\begin{itemize}
\item Removing data points symmetrically from the minimum of the correlator;
\item Adding data points symmetrically at the minimum of the correlator, by
  \begin{itemize}
  \item Adding zeroes;
    \item Adding the minimum value of the correlator.
  \end{itemize}
\end{itemize}
No significant difference is observed between these methods for both real and synthetic test data. Since the correlator is exponentially suppressed in the region where data points are added or removed, this is not unexpected. This is in contrast to the case of light mesons, where artefacts due to padding with zeroes can be observed \cite{Quinn:2019uwq}, and more sophisticated methods must be applied \cite{Quinn:2019mp468p}.  Padding with data points can be done at all temperatures in contrast to removing points where $N_{0} > 3N_{\tau}$ is required.

Next we compared the ratio of the actual and the reconstructed correlator with the double ratio introduced in \Refltwo{Aarts:2022krz}{Aarts:2023nax}. For the latter, no padding or subtraction is required. We found that both methods give comparable insight into the possible temperature of the spectral content in the correlator. 


\begin{acknowledgments}
GA, CA, RB and TJB are grateful for support via STFC grant ST/T000813/1. 
MNA acknowledges support from The Royal Society Newton International Fellowship. RB acknowledges support from a Science Foundation Ireland Frontiers for the Future Project award with grant number SFI-21/FFP-P/10186. This work used the DiRAC Extreme Scaling service at the University of Edinburgh, operated by the Edinburgh Parallel Computing Centre and the DiRAC Data Intensive service operated by the University of Leicester IT Services on behalf of the STFC DiRAC HPC Facility (www.dirac.ac.uk). This equipment was funded by BEIS capital funding via STFC capital grants ST/R00238X/1, ST/K000373/1 and ST/R002363/1 and STFC DiRAC Operations grants ST/R001006/1 and ST/R001014/1. DiRAC is part of the UK National e-Infrastructure. We acknowledge the support of the Swansea Academy for Advanced Computing, the Supercomputing Wales project, which is part-funded by the European Regional Development Fund (ERDF) via Welsh Government, and the University of Southern Denmark and ICHEC, Ireland for use of computing facilities. This work was performed using PRACE resources at Cineca (Italy), CEA (France) and Stuttgart (Germany) via grants 2015133079, 2018194714, 2019214714 and 2020214714.
\end{acknowledgments}

\section*{Software and Data}

Correlation functions were generated using \textsc{openQCD-hadspec}~\cite{glesaaen_jonas_2018_2217028}, an extension to \textsc{openQCD-FASTSUM}~\cite{glesaaen_jonas_rylund_2018_2216356} for correlation functions. \textsc{openQCD-FASTSUM} was used for ensemble generation~\cite{Aarts:2020vyb,Aarts:2022krz}.
The analysis in this work, available at \Refl{ZenodoLattice23Bignell}, makes extensive use of the \textsc{python} package \textsc{gvar}~\cite{peter_lepage_2022_7315961}. Additional data analysis tools included \textsc{matplotlib}~\cite{Hunter:2007,thomas_a_caswell_2022_6513224}, \textsc{numpy}~\cite{harris2020array} and \textsc{lsqfit}~\cite{peter_lepage_2021_5777652}. The zero temperature mass results of \Refl{Aarts:2023nax} were used in this study. Nucleon mass fits used the scripts available at \Refl{aarts_gert_2023_8273591} with an input file available at \Refl{ZenodoLattice23Bignell}.

\section*{Authors' Contributions}

Bignell: Data analysis, plot generation and primary manuscript production. 
J\"ager: Data production.
Aarts, Allton, Anwar, Burns, Skullerud: Contributions to analysis and physics interpretations of results, and draft of the manuscript.

\subsection*{Open Access Statement}
For the purposes of open access, the authors have applied a Creative Commons Attribution (CC BY) to any Author Accepted Manuscript version arising.
%
\bibliographystyle{JHEP}
\bibliography{skeleton}

\end{document}